\documentclass[12pt]{article}
\usepackage{amstex,amssymb}

\newcommand{\re}{{\rm I\!\rm R}}        

\numberwithin{equation}{section}

\begin{document}

\begin{flushright}
SU-4240-672 \\
\today
\end{flushright}

\begin{center}
{\large{\bf SKYRMIONS, SPECTRAL FLOW AND PARITY DOUBLES}}

\bigskip
   A. P. Balachandran,\,
   and \, Sachindeo Vaidya\footnote{sachin@@suhep.syr.edu} \\ 
   {\it Department of Physics, Syracuse University,} \\ 
   {\it Syracuse, N . Y.  13244-1130, U. S. A.}
\end{center}

\begin{abstract}
It is well-known that the winding number of the Skyrmion can be identified
as the baryon number. We show in this paper that this result can also be
established using the Atiyah-Singer index theorem and spectral flow
arguments. We argue that this proof suggests that there are light quarks
moving in the field of the Skyrmion. We then show that if these light
degrees of freedom are averaged out, the low energy excitations of the
Skyrmion are in fact spinorial. A natural consequence of our approach is
the prediction of a $\frac{1}{2}^{-}$ state and its excitations in addition
to the nucleon and delta. Using the recent numerical evidence for the
existence of Skyrmions with discrete spatial symmetries, we further suggest
that the the low energy spectrum of many light nuclei may possess a parity
doublet structure arising from a subtle topological interaction between the
slow Skyrmion and the fast quarks. We also present tentative experimental
evidence supporting our arguments.
\end{abstract}

\section{Introduction}
The modern understanding of high energy physics is based on the Standard
Model, the strong interaction sector of which is described by QCD. It is
however difficult to make reliable statements about the low energy
behavior of QCD where its coupling constant becomes large and perturbation
theory breaks down. Several phenomenological models have been developed in
an attempt to understand its low energy dynamics, like for example the
chiral model (see for example \cite{bhaduri} and references therein). In
this model, the chiral field $U$, which is valued in $SU(N_f)$ for $N_f$
flavors, describes the low energy excitations about the QCD ground
state. The baryon octet can be identified with the solitons of the Skyrme
Lagrangian \cite{skyrme}, the baryon number $B$ being the winding number
of the soliton \cite{bnrs1,bnrs2,witten1,witten2}.

In the original research, static solutions of the Skyrme Lagrangian
\cite{skyrme} for the winding number $\pm 1$ sectors were found and these
solitons used for the description of $|B| = 1$ baryons. Since that work,
there has been a considerable amount of interest in searching for Skyrme
solitons with higher winding numbers as well. In particular, a $|B|=2$
solution appropriate to describe the H-dibaryon
\cite{jaffe1,jaffe2,bblrs,blrs,jafkor,yosnap} and also numerical
solutions with the symmetry of discrete sub-groups of $SO(3)$
\cite{brtoca,carson,batsut} have been found.

In this paper, we largely restrict ourselves to two flavors. A standard
parity-invariant manner in which the quarks couple to the background
$U$-field is via the Yukawa coupling,
\begin{eqnarray}
{\cal L}_{int} &=& -m[\overline{\psi}_LU^{\dagger} \psi_R +
		   \overline{\psi}_R U \psi_L]  \label{ly1} \\
	&\equiv& -m\overline{\psi} e^{-i \gamma_5 \vec{\tau}. \vec{\phi}}
		   \psi, \\
	U &=& e^{i \vec{\tau}.\vec{\phi}}, \text{$\tau_i$=Pauli matrices}
\end{eqnarray}
where $\psi_R = \frac{1+\gamma_5}{2} \psi$, $\psi_L =
\frac{1-\gamma_5}{2}\psi$. The different winding number sectors of this
$U$-field are precisely the Skyrmions. There are of course no histories of
$U$ that can change its winding number. We will consider a generalization
of this Lagrangian that will effectively allow us to study winding number
changing histories. Using the index theorem and the associated spectral
flow picture, we will show that the one-fermion state in the background of
the soliton corresponds to the fermionic vacuum with no soliton. This leads
to a natural interpretation for the baryon number of the Skyrme
soliton. Specifically, we will show that the nucleon (with quantum numbers
$I (J^P) = \frac{1}{2} (\frac{1}{2}^+)$ is described by taking the even
combination of this fermionic state and its parity transform. The
parity-odd combination describes some excited states with $I (J^P) =
\frac{1}{2} (\frac{1}{2}^-)$, perhaps the $N(1535)$ or $N(1650)$.

This argument also leads to two interesting consequences. First of all,
since there are quarks created in the process of creating a Skyrmion, they
may stick to the Skyrmion. Thus one can think of the baryon as made up of
not just the ``soliton lump'', but also of these fast quarks whizzing
around it. As these quark degrees of freedom are of considerably higher
energies than those of the collective motion of the soliton, the low energy
analysis of the baryon can be based on the Born-Oppenheimer (B-O)
approximation. If these quark degrees of freedom are now averaged out, the
low energy excitations of the Skyrmion become spinorial. This is a novel
argument to show that the Skyrmion is a fermion. (A related argument,
demonstrating the effect on the Dirac sea of a $2\pi$-rotation of the
Skyrmion, has been discussed by \cite{gomez}.)

The second interesting consequence arises if one considers Skyrmions with
discrete symmetries. The general theory for doing the B-O approximation for
systems with discrete symmetries was discussed by two of the authors in
\cite{balvai1,balvai2}. It was shown that for such systems, the B-O
approximation leads to quite subtle quantum effects. In particular, in
previous work \cite{basiwi} and here, it has been argued that many shapes
lead to anomalous violation of parity in quantum theory if care is not
taken to judiciously include aspects of microscopic degrees of freedom as
well. On including these high energy degrees of freedom, it was argued that
the low energy excitation spectrum of the system should display a ``parity
doublet structure''. An example of such a doublet is the pair of closely
spaced levels of the ammonia molecule used in the construction of the
ammonia maser. It was suggested in \cite{balvai1} that such doubles should
occur among bound states involving quarks as well. We now argue that parity
doubles of this kind occur in Skyrmion physics too and make specific
phenomenological predictions. Parity doubles among light baryons have also
been discussed by \cite{iachello,kirchbach}.

The organization of this article is as follows: In Section 2, using the
appropriate index theorem, we show that as the Skyrmion is created, quarks
are created as well. We then argue in Section 3 that if these fast quark
degrees of freedom are averaged out, the low energy excitations of the
Skyrmion become in fact spinorial. In Section 4, we consider Skyrmions with
discrete symmetry groups, and suggest that some of these systems possess a
parity doublet structure for their low energy spectrum. We also present
tentative experimental evidence for the existence of parity doublets in the
low energy excitation spectrum of several light nuclei. Section 5 is
devoted to concluding remarks.

\section{Spectral Flow and Particle Creation}
Let us consider the two-flavored chiral model, so that $U(\vec{x}, t) \in
SU(2)$. The Skyrme Lagrangian is
\begin{equation}
{\cal L}_{Skyrme} = \frac{f^2_{\pi}}{16}{\rm Tr} U^{\dagger}\partial_{\mu}U
U^{\dagger} \partial^{\mu}U + \frac{1}{32e^2}{\rm
Tr}[U^{\dagger}\partial_{\mu}U, U^{\dagger} \partial_{\nu}U]^2.
\label{lskyrme}
\end{equation}

A standard parity invariant interaction term for an $SU(2)$ doublet of
Dirac fermions in the background of the chiral field $U$ is of the form
(\ref{ly1}). We can identify $\psi$ with the quark fields $q^{\alpha}$
after attaching a color index $\alpha$.  

Let us now redefine the fields $\psi_L$ and $\psi_R$ as follows:
\begin{eqnarray}
\Psi_L &=& \psi_L, \nonumber \\ 
\Psi_R &=& U^{\dagger} \psi_R.
\label{redefs}
\end{eqnarray}
The fermion Lagrangian density (ignoring the Skyrme Lagrangian for the
moment) then becomes
\begin{equation}
{\cal L}_F = \overline{\Psi}_L i\gamma^{\mu} \partial_\mu \Psi_L
+\overline{\Psi}_R i\gamma^{\mu} ( \partial_\mu +U^{\dagger} \partial_\mu U)
\Psi_R - m[\overline{\Psi}_R \Psi_L + {\rm h.c.}].
\label{ly2}
\end{equation}

We can treat the mass term here as a perturbation and thus initially ignore
it. Therefore, to zeroth order in $m$, the Lagrangian of interest is
\begin{equation}
{\cal L}_F = \overline{\Psi}_L i\gamma^{\mu} \partial_\mu \Psi_L + 
 \overline{\Psi}_R i\gamma^{\mu} ( \partial_\mu +U^{\dagger} \partial_\mu U)
\Psi_R.
\label{lu}
\end{equation}

For fixed time, $U$ is a map from $\re^3$ into $SU(2)$, with the condition
that $U(x) \rightarrow {\bf 1}$ as $|\vec{x}| \rightarrow \infty$ $(x
\equiv \vec{x}, t)$, that is, $U \in {\rm Maps}(S^3 \rightarrow SU(2))$.

The space of $U$'s is disconnected, since
\begin{equation}
\pi_0 [{\rm Maps}(S^3\rightarrow SU(2))] = \pi_3[SU(2)] = {\mathbb Z}.
\end{equation}
The different disconnected pieces or sectors are labeled by the ``winding
number'' $n$ of $U$:
\begin{equation}
n = \frac{1}{24 \pi^2} \int Tr(U^{\dagger}dU)^3 .
\label{solnum}
\end{equation}
Here we are using the notation of differential forms, and the wedge product
between differential forms is being omitted.

Clearly, $U$-fields with different winding numbers give rise to different
quantum fermionic systems via (\ref{ly2}). We are interested in
understanding the effect on the fermionic system as the winding number is
changed.  Specifically, we would like to follow the energy levels of the
quantum fermionic Hamiltonian as the winding number is changed. In order to do
this, let us consider a slightly more general Lagrangian
\begin{equation}
{\cal L}' = \overline{\Psi}_L i\gamma^{\mu} \partial_\mu \Psi_L + 
\overline{\Psi}_R i\gamma^{\mu} (\partial_\mu + A^R_{\mu}) \Psi_R.
\label{la}
\end{equation}
This Lagrangian is invariant under gauge transformations of the form
\begin{eqnarray}
\Psi_R &\rightarrow& g^{-1}\Psi_R, \\
A^R_{\mu} &\rightarrow& g^{-1}A^R_{\mu}g+g^{-1} \partial_\mu g, \\
\Psi_L &\rightarrow& \Psi_L, \\
g(x) &\in& SU(2).
\end{eqnarray}
Had we added the kinetic energy term
\begin{equation}
\text{constant}\times \int {\rm Tr}[F^{\mu \nu R} F^R_{\mu \nu}] \nonumber  
\end{equation}
where $F^R_{\mu \nu}$ is the curvature of $A^R_{\mu}$, we would
have a chiral $SU(2)$ gauge theory.

For flat connections, (\ref{la}) reduces to (\ref{lu}). 

The advantage of using the more general Lagrangian (\ref{la}) is that there
is no need to restrict to flat connections. For this reason, we have the
possibility of studying winding number changing histories as we now show.
Note that such histories do not exist in the space of $U$'s.

With this more general Lagrangian, we can now imagine continuous paths (in
the configuration space of the gauge field $A^R_\mu$) connecting the two
flat connections $A^R_\mu=0$ and $A^R_\mu = U^{\dagger} \partial_\mu U$. If
$U$ has a non-trivial winding number $n$, then these paths necessarily go
through configurations having non-trivial gauge field strengths. If we
choose instanton-like interpolating fields, the spectral flow of the
relevant Dirac Hamiltonian can be computed, as we shall see.

Let us consider a static Skyrme field $U, U(x)=U_c(\vec{x})=e^{i
\theta_c(r) \vec{\tau}.\hat{x}}=\cos \theta_c(r) + i \vec{\tau}.\hat{x}
\sin \theta_c(r)$, of winding number $n=1$, where $r=|\vec{x}|$ and
$\hat{x}=\vec{x}/|\vec{x}|$.  While there are no continuous paths that
interpolate between $U$'s of different winding numbers, we can find
continuous paths interpolating between any two connections. Let us thus
look at instanton-like configurations of the gauge field $A^R$ (in the
temporal gauge $A^R_0=0$) of instanton charge $q_R \equiv 1/16\pi^2 \int
Tr(F^R \wedge F^R) = -1$, connecting $A^R_i=0$ to $A^R_i=U_c^{\dagger}
\partial_i U_c$. Since the winding number of a field $g$ is defined as
$1/{24\pi^2} \int Tr(g^{-1}dg)^3$ [cf. (\ref{solnum})], we can see that the
instanton charge $q_R$ is simply the difference between winding numbers of
the fields ${\bf 1}$ and $U_c$. .

Let us look at the 4-dimensional (Euclidean) Dirac equation in the presence
of such an instanton field,
\begin{equation}
\left[ \begin{array}{cc}
	 0 & \tilde{L} \\
	 L & 0 
	\end{array} \right]  \left[ \begin{array}{c}
					\chi_R \\
					\chi_L
				    \end{array} \right] = 0,
\label{de}
\end{equation}
where 
\begin{equation}
\Psi_R = \left[ \begin{array}{c}
		\chi_R \\
		 0
		\end{array} \right], \Psi_L = \left[ \begin{array}{c}
							0 \\
		 				      \chi_L
						     \end{array} \right]
\end{equation}
in the basis in which $\gamma_5$ is diagonal. Also $L$ and $\tilde{L}$ are
defined as
\begin{equation}
L = i \bar{\alpha}^{\mu}(\partial_{\mu} + A^R_{\mu}), \quad \tilde{L} = i
	\alpha_{\mu} \partial_{\mu}
\end{equation}
where $\alpha^{\mu} = (-i \vec{\sigma}, {\bf 1})$ and $\bar{\alpha}^{\mu} =
(i \vec{\sigma}, {\bf 1})$. We are using Dirac matrices of the form
\begin{equation}
{\gamma}^{\mu} = \left[ \begin{array}{cc}
                            0        &   {\alpha}^{\mu} \\
                         \bar{\alpha}^{\mu} &   0
                        \end{array} \right].     \nonumber
\end{equation} 
With this choice of representation, ${\gamma}^5 = {\gamma}^1
{\gamma}^2 {\gamma}^3 {\gamma}^4 = \text{diag}({\bf 1}, -{\bf 1})$ is
diagonal. The equation (\ref{de}) is a consequence of (\ref{la}).

Let us look for non-trivial solutions, or the zero modes of the equations
\begin{eqnarray}
L \chi^R = 0, \label{eR}\\
\tilde{L} \chi^L = 0. \label{eL}
\end{eqnarray}
Multiplying (\ref{eR}) by $L^{\dagger}$ and (\ref{eL}) by
$\tilde{L}^{\dagger}$, we obtain 
\begin{eqnarray}
((\partial_{\mu} + A^R_{\mu})^2 + 2i \bar{\sigma}^{\mu \nu} F_{R\mu \nu})
\chi^R &=& 0, \label{e+} \\        
{\bf 1}(\partial_{\mu} \partial_{\mu}) \chi^L &=& 0, \label{e-}
\end{eqnarray}
where $\bar{\sigma}^{\mu \nu}=(1/4i)({\alpha}^{\mu} \bar{\alpha}^{\nu} -
{\alpha}^{\nu} \bar{\alpha}^{\mu})$. In general, it is difficult to know
the number of solutions of (\ref{e+}). However, following \cite{janore} we
can settle this number if we use an anti self-dual instanton field for
$A^R$ (that is, $F^R_{\mu \nu} = -1/2 \epsilon_{\mu \nu \rho \lambda} F^{R
\rho \lambda}$) (and it is always possible to find one
\cite{atiman}). Since (\ref{e-}) has only the trivial solution $\chi^L =0$,
the Atiyah-Singer index theorem tells us that the number of linearly
independent solutions of (\ref{e+}) is exactly $|q_R|$, which is 1 in our
case.

Following Witten \cite{witten}, let us show that the number of zero modes
of (\ref{eR}) is in fact equal to the spectral flow of the
Dirac Hamiltonian 
\begin{equation}
H_R = -i\sigma^k(\partial_k + A^R_k) \label{3diracR}
\end{equation}
for $\chi_R$. Rewriting (\ref{eR}) as
\begin{equation}
\partial_4 \chi^R=-i\sigma^k(\partial_k + A^R_k) \chi^R,
\label{de0}
\end{equation}
let us adiabatically change the gauge field $A^R(\vec{x}, x_4)$ as the
Euclidean time $x_4$
changes from $-\infty$ ( where $A^R =U^{\dagger}_cdU_c$) to $+\infty$
(where $A^R =0$). For $\chi^R$, let us make the ansatz
\begin{equation}
\chi^R(\vec{x}, x_4) = F(x_4) \phi^R_{x_4}(\vec{x}),
\end{equation}
where $\phi^R_{x_4}(\vec{x})$ is an eigenfunction of (\ref{3diracR}):
\begin{equation}
H_R \phi^R_{x_4}(\vec{x}) = \lambda(x_4)\phi^R_{x_4}(\vec{x}).
\end{equation}

Substituting the above ansatz for the zero mode of (\ref{eR}) in
(\ref{de0}), we get
\begin{equation}
F(x_4) = F(0) \exp \left( \int^{x_4}_0 d \tau' \lambda(\tau') \right).
\end{equation}
It is obvious that $\chi^R(\vec{x}, x_4)$ is normalizable only if $\lambda$
is negative for $\tau \rightarrow \infty$ and positive for $\tau
\rightarrow -\infty$.  Thus the existence of the zero mode of (\ref{eR})
necessarily implies that one right-handed fermionic level of the
$H_R$ flows from positive to negative value. In other words, we find that
the 1-particle state in the background field $U_c$ gets mapped to the
vacuum state of zero winding number. We can therefore unambiguously say
that fermionic vacuum in the background of the trivial winding number is
the single particle state in the background field of winding number one.

We need to remark on an important issue of the quantum numbers of this
fermionic state. First of all, the QCD Lagrangian (and its effective low
energy description) is invariant under parity. So, under a parity
transformation of the above right-handed fermionic state, we get a
left-handed state as well. The correct description of the nucleon in the
ground state is via the parity-even combination of these two states. The
parity-odd combination corresponds to excited states with quantum numbers
$I (J^P) = \frac{1}{2} (\frac{1}{2}^-)$. Secondly, QCD is a theory with
$N_c = 3$ colors. So on second quantization, the above state must be
populated by three quarks forming a color singlet.

Let us take stock of the implication of such a spectral flow for our
situation. Recall that we started from the Lagrangian (\ref{lu}), which we
generalized to (\ref{la}). Using the instantons available in this more
general Lagrangian, we found that we could study paths connecting winding
numbers 0 and 1. In terms of the original Lagrangian (\ref{lu}), what this
means is that one can relate the fermionic spectrum for a winding number
zero background $U$-field to that for a winding number 1 $U$-field. In
particular, there is a history which maps the vacuum of the first situation
to the one-particle state of the second.

Higher winding number Skyrmions can be similarly discussed by using
instantons of higher instanton number $|q|$. This results in $|q|$ energy
levels flowing from negative to positive values.

Thus we can unambiguously say that processes that change the winding number
of the $U$-field by $n$ are accompanied by the ``binding'' of $n$ Dirac
fermions to the soliton. The winding number can thus be interpreted as the
fermion (or baryon) number.

\section{Skyrmion Spin from B-O approximation}
The spectral flow argument presented in the previous section strongly
suggests that there are quarks moving in the background of the Skyrme field
$U$. The system thus naturally lends itself to a separation into
the ``fast'' quark degrees of freedom, and the ``slow'' Skyrme degrees of
freedom. It is therefore plausible that the simplest attempt to quantize this
system would be via the Born-Oppenheimer approximation or some version
thereof. 

The bundle-theoretical formulation of B-O approximation has been reviewed
and elaborated in \cite{moshwi,balvai2}. Let us recall it briefly. The B-O
approximation is used to describe systems whose Hamiltonian ${\cal H}$
conveniently splits into ${\cal H}_f + {\cal H}_s$, where we can roughly
think of ${\cal H}_f$ as dictating the ``fast'' dynamics, and ${\cal H}_s$
the ``slow'' dynamics. Let us {\it assume} that this Hamiltonian ${\cal H}$
has eigenfunctions that are sections of a trivial bundle over the slow
configuration space $Q_{slow}$. [ We can modify this hypothesis and hence
the subsequent discussion by substituting twisted for trivial bundles, but
this change is not appropriate for Skyrmions.] The fast wave function
$\psi_f$ is an eigenfunction of ${\cal H}_f$ (or more generally some linear
combination of the eigenfunctions of ${\cal H}_f$) and is a function on
both the fast configuration space $Q_{fast}$ and the slow configuration
space $Q_{slow}$. In particular, the fast wave function is the section of
some bundle (generally twisted) over $Q_{slow}$. The slow wave function is
calculated not from ${\cal H}_s$ but by averaging ${\cal H}_s$ over
$\psi_f$ to give the emergent ``slow'' B-O Hamiltonian $\hat{\cal
H}_s$. This Hamiltonian generally contains a connection, and eigenfunctions
of $\hat{\cal H}_s$ are also sections of a (generally) twisted bundle over
$Q_{slow}$ determined by this connection. But it has a twist exactly
``opposite'' to that of $\psi_f$ in the sense that the product
$\psi_f. \psi_s$ can be projected onto the trivial bundle over $Q_{slow}$.

Let us elaborate on the system under consideration in order to apply the
above ideas. We have a $U$-field that describes a soliton and has some
non-trivial winding number, and there are some quarks (the number of quarks
being given by the topological arguments of the previous section, and by
color confinement) moving in the vicinity of this soliton. The quarks are
to be quantized with this $U$-field in the background. (Quark wave
functions in the background of a Skyrmion have been studied by
\cite{kahrip,kariso} in a slightly different context.) For a complete
quantum-mechanical description of low energy physics, we must now
quantize the collective motion of the Skyrmion as well. We will demonstrate
now that in the B-O approximation, when the fast quarks are averaged out,
the low energy excitations of the Skyrmion are spinorial.

For simplicity, let us begin with the hedgehog configuration $U_c(\vec{x})
= \cos \theta(r) {\bf 1} + i \vec{\tau}. \hat{x}\; \sin \theta(r)$. Under
an isospin transformation by $A \in SU(2)$, we get in general a different
configuration $AU_c(\vec{x})A^{\dagger}$, but with the same energy. (Note
that since a transformation by $A$ or $-A$ give the same configuration, the
space of these configurations is actually diffeomorphic to $SU(2)/{\mathbb
Z}_2=SO(3)$.)  Similarly, under a non-trivial spatial rotation $\vec{x}
\rightarrow R \vec{x}$, we get another configuration $U_c(R \vec{x})$ that
is distinct from $U_c(\vec{x})$, but still with the same energy. We call
all these configurations that are distinct from $U_c$ but with the same
energy as the ``collective or zero modes''. It would seem that the space of
zero modes is $SO(3) \times SO(3)$, but this is not so. This is because of
the identity $A\tau_i A^{\dagger}=\tau_jR_{ji}(A)$, which means that every
spatial rotation is equivalent to some isospin transformation. The space of
zero modes is thus topologically $SO(3)$.

We will occasionally refer to the space of $A$'s, which is topologically
$S^3 \approx SU(2)$ as the space of zero modes although the latter is in
reality obtained from $SU(2)$ only after identifying $A$ with $-A$.

According to the standard prescription of collective coordinate
quantization, we can assume that $U$ is given by
\begin{equation}
U(\vec{x}, t) = A(t)U_c(\vec{x})A(t)^{\dagger}, \quad A(t) \in SU(2).
\label{ansatz}
\end{equation}
The configuration space $Q_{slow}$ of the $U$'s can be identified with
that of the collective modes. 

How does $A$ transform under spatial and isospin rotations? Recall that
under a spatial rotation, the hedgehog ansatz $U_c(\vec{x})$ changes to
$U_c(R \vec{x}) = g_R U_c(\vec{x}) g^{\dagger}_R$, where $R$ is determined
by $g_R$ via the relation $g_R \tau_i g^{\dagger}_R = \tau_j R_{ji}$. This
is equivalent to the transformation $A \rightarrow Ag_R$ by
(\ref{ansatz}). Similarly, under an isospin transformation by $g_I$,
$U(\vec{x}) \rightarrow g^{\dagger}_I U(\vec{x}) g_I$. This is equivalent
to the transformation $A \rightarrow g^{\dagger}_I A$. Thus spatial
rotations act on the spaces of $A$'s by {\it right} multiplication, while
isospin rotations act by {\it left} multiplication \cite{BMSS}. The
transformations on the space of $A$'s are therefore done by operators
${\cal R}_s$ and ${\cal I}_s$ as follows:
\begin{eqnarray}
{\cal R}_s(g_R)A{\cal R}^{\dagger}_s(g_R) &=& Ag_R, \nonumber \\
{\cal I}_s(g_I)A{\cal I}^{\dagger}_s(g_I) &=& g^{\dagger}_I A.
\end{eqnarray}
The subscript $s$ for ${\cal R}_s$ and ${\cal I}_s$ is to indicate that
these operators act on the slow degrees of freedom.

Let us describe the ``fast'' wave function $\psi_f$ in a little more
detail. This $\psi_f$ is an eigenfunction of the Dirac Hamiltonian, with
the connections $A^L_i = 0$ and $A^R_i = U^{\dagger} \partial_i U$. The
ansatz $U$ is invariant under combined spatial and isospin rotations, but
not under separate spatial rotations or separate isospin
transformations. This in turn implies that the symmetry group for the
Hamiltonian ${\cal H}_f$ of $\psi_f$ is generated only combined spatial and
isospin rotations. Thus we can arrange for $\psi_f$ to be an eigenfunction
of $\vec{K} \equiv \vec{I}_f + \vec{J}_f$, where $\vec{I}_f$ is the isospin
and $\vec{J}_f$ the total angular momentum of $\psi_f$. [These do not
include isospin and angular momentum of the $U$-field.] The total angular
momentum $\vec{J}_f$ of $\psi_f$ is $\vec{S}_f+\vec{L}_f$, $\vec{S}_f$
being the spin and $\vec{L}_f$ the orbital angular momentum. Now,
$\vec{K}^2$ has eigenvalues $k(k+1)$ with $k= 0, 1/2, 1..$. The eigenvalue
$k(k+1)$ is $(2k+1)$-fold degenerate. However, in the previous section, we
argued that when a Skyrmion of winding number 1 is ``created'', a single
fermion state is also created. Thus, consistency with the spectral flow
picture requires us to choose $\vec{K}=0$, as that is the only possibility
without degeneracy. But this in turn implies that $\vec{J}_f=1/2$. Now, a
$\vec{J}_f=1/2$ state can be created by combining $\vec{S}_f$ with
$\vec{L}_f = $ 0 or 1. The correct even parity of the nucleon is reproduced
only by choosing $\vec{L}_f=0$, since parity is given by
$(-1)^{L_f}$. States with non-zero $L_f$ would have higher energy because
of the centrifugal barrier, so choosing $\vec{L}_f = 1$ would provide the
description of the excited states $I (J^P) = \frac{1}{2} (\frac{1}{2}^-)$.

Let us examine what $\psi_f$ looks like for $U=U_c$, that is, for $A = {\bf
1}$. For our arguments, it is only the spin and isospin dependences of
$\psi_f$ that matters. Purely from symmetry considerations, we can find out
these dependences. 

When $\vec{K}=0$, $\psi_f$ is of the form
\begin{equation}
\psi_f:=\psi_f({\bf 1})= \left( \begin{array}{c}
                                   \psi_f^{+}({\bf 1}) \\
                                   \psi_f^{-}({\bf 1}) 
                                  \end{array} \right), \quad \label{fwave}
\end{equation}
where  
\begin{equation}
\psi^{\pm}_f({\bf 1})=(\psi^{\pm}_f({\bf 1})_{\alpha \beta}), 
\end{equation}                                     
and $\alpha, \beta$ are spin and isospin indices respectively:         
\begin{equation}
(\psi^{+}_f({\bf 1})_{\alpha \beta}) = \epsilon_{\alpha \beta}
                               \eta^{+}_f({\bf 1}),
\end{equation}
where
\begin{eqnarray}
\eta^{+}_f({\bf 1}) &=& |S=1/2, S_z = 1/2> |I=1/2, I_z=-1/2> \nonumber \\
	       &\; &- |S=1/2, S_z =-1/2> |I=1/2,I_z=1/2>  \nonumber \\
	        &\equiv& \epsilon^{ab}|1/2, S_a>|1/2, I_b>, \label{eta}
\end{eqnarray} 
where $a, b =1, 2$ and $S_1 = I_1 = 1/2, S_2 = I_2 = -1/2$. The wave
function $\eta^{+}_f({\bf 1})$ has the desired behavior under combined
spin-isospin transformations. (We have suppressed the $\vec{x}$-dependence
of this wave function since it is unnecessary for our arguments.) The state
$\psi^{-}_f({\bf 1})$ is similarly defined. Note that the wave function
(\ref{fwave}) must be {\it even} under parity, and so $\eta^{-}_f({\bf 1})
= - \eta^{+}_f({\bf 1})$.

The wave function $\psi_f(A)$ for arbitrary $A$ follows from (\ref{eta}) by
isospin rotation and is given by 
\begin{equation}
\psi_f(A) =\left(\begin{array}{c}
                    \psi_f^{+}(A) \\
                    \psi_f^{-}(A) 
                   \end{array} \right), \label{fastA}
\end{equation}
where
\begin{eqnarray}  
(\psi^{+}_f(A))_{\alpha \beta} &=& \epsilon_{\alpha \beta}\eta^{+}_f(A), \\ 
\eta^{+}_f(A) &=& \epsilon^{ab}|1/2, S_a>|1/2, I_c>A_{cb}.
\end{eqnarray}
There are also similar expressions for $\psi^{-}_f(A)$ and
$\eta^{-}_f(A)$. 

The transformation properties of $\psi_f(A)$ are determined by those of
$\eta^{\pm}_f(A)$.

Under spatial and isospin transformations, $\eta^{+}_f({\bf 1})$ transforms
as
\begin{eqnarray}
{\cal R}_f(g_R) \eta^{+}_f({\bf 1}) &=& \epsilon^{ab}|1/2, S_c>|1/2, I_b>
(g_R)_{ca}, \label{etarotate} \\ 
{\cal I}_f(g_I) \eta^{+}_f({\bf 1}) &=& \epsilon^{ab}|1/2, S_a>|1/2,
I_c> (g_I)_{cb},
\end{eqnarray}
where the subscript $f$ for ${\cal R}_f$ and ${\cal I}_f$ is to indicate
that these operators act on the fast degrees of freedom.

We will show that $\psi_f(A)$ is {\it invariant} under the action of {\it
total} spatial rotations as well as {\it total} isospin rotations, where
these act on both $\psi_f$ and on the Skyrmion zero mode $A$. In other
words, if we rotate the Skyrmion (by the action of rotations on $A$) and at
the same time rotate the functions $\eta^{\pm}_f(A)$ using
(\ref{etarotate}), we find that $\psi_f(A)$ is invariant
\cite{balvai2}. The same goes for isospin rotations as well.

The proof is as follows. Under total rotation by $g_R$,
\begin{eqnarray}
\eta^{+}_f(A) &\rightarrow& {\cal R}_s(g_R)A{\cal R}_s(g^{\dagger}_R) 
             {\cal R}_f(g_R) \eta^{+}_f ({\bf 1}) \nonumber \\
   &=&\epsilon^{ab}|1/2, S_c>|1/2, I_d> (g_R)_{ca} (Ag_R)_{cb}, \nonumber \\
   &=&\epsilon^{ab}|1/2, S_a>|1/2, I_d>A_{db}, \nonumber \\
   &=&\eta^{+}_f(A) \label{rotationeta}.
\end{eqnarray}
Similarly, under total isorotation by $g_I$,
\begin{eqnarray}
\eta^{+}_f(A) &\rightarrow& {\cal I}_s(g_I)A{\cal I}^{\dagger}_s(g_I) {\cal
I}_f(g_I)\eta^{+}_f ({\bf 1}) \nonumber \\
   &=&\epsilon^{ab}|1/2, S_a>|1/2, I_c> (g_I)_{cd}(g^{\dagger}_I A)_{db},
\nonumber \\
   &=& \epsilon^{ab}|1/2, S_a>|1/2, I_c> A_{cb}, \nonumber \\
   &=& \eta^{+}_f(A) \label{isospineta}.
\end{eqnarray}

Similarly, $\eta^{-}_f(A)$ is invariant under total rotations and total
isorotations as well.
   
The wave function $\psi_f(A)$ defines the section of a bundle twisted over
$Q_{slow}$. This follows from the definition (\ref{fastA}) of
$\psi_f(A)$ which shows that 
\begin{equation}
\psi_f(-A) = -\psi_f(A).
\end{equation}
 
Let us now discuss the wave function of the slow part, that is, of the
collective modes of the Skyrmion. On substituting the ansatz (\ref{ansatz})
in the Skyrme Lagrangian (\ref{lskyrme}), we can easily find the Lagrangian
for the ``zero-modes'' $A(t)$,
\begin{equation}
L_{Skyrme} = \int d^3 x {\cal L}_{Skyrme} = -\frac{1}{2}a(U_c){\rm
Tr}(A^{\dagger} \dot{A})^2 - E(U_c),
\end{equation}
where $a(U_c)$ and $E(U_c)$ are positive constants (see \cite{BMSS} for
details). 
The Hamiltonian calculated from above is 
\begin{equation}
{\cal H}_{slow} = \frac{2}{a(U_c)} L_{\alpha} L_{\alpha} + E(U_c),
\end{equation}
where $L_{\alpha}$ are the ``left'' rotation generators acting on sections
of the trivial bundle over $Q_{slow}$. They consist of even functions of
$A$ and carry integer angular momenta.

However, as emphasized earlier, ${\cal H}_{slow}$ is not the B-O
Hamiltonian. The correct B-O Hamiltonian is obtained by averaging the
${\cal H}_{slow}$ over $\psi_f(A)$. By the general arguments presented, we
know that this averaging produces a connection, which can affect the
bundle-theoretic character of its eigenstates.

This bundle is actually twisted, as can be seen by parallel transporting
the {\it full} wave function $\psi_s.\psi_f$ in a closed non-contractible
loop, defined by $ \{ Ae^{i \tau_3 \alpha/2}, \alpha \in [0, 2\pi] \}$.  By
hypothesis, the full wave function is unchanged under this transport.
Therefore, since $\psi_f$ changes its sign, so must $\psi_s$. Transporting
$\psi_f(A)$ in this closed loop cannot be interpreted as a sequence of
isospin transformations from 0 to $2 \pi$. But as far as $\psi_s$ is
concerned, this transport in the closed loop is in fact such a sequence of
isospin transformations on the slow variables $A$, generated by its isospin
operators. The slow wave function is thus an isospinor. Similar arguments
show that it is a spinor under spatial rotations as well. The total wave
function in view of (\ref{rotationeta}) and (\ref{isospineta}) thus
transforms as a spinor under spatial and isospin rotations separately.

A very interesting picture emerges from the above discussion. On the
one hand, the Skyrmion can be thought of the classical lump $U$ with quarks
sticking to it, as suggested by the spectral flow. This composite has the
correct quantum numbers of a nucleon. (with $N_c = 3$). A dual description
can also be constructed, in which the quarks are averaged out. In this
picture, the slow wave function is seen to be spinorial. The two
descriptions are in fact the two ``limits'' of the unified description
involving both the fast and the slow degrees of freedom. In the unified
picture, the Skyrmion is only a part of the full description. The full wave
function $\psi_{total} = \psi_f(A,\vec{x}). \psi_s(A)$ is also spinorial,
and thus describes the nucleon.

\section{Parity Doubles in Light Nuclei}
\subsection{Non-spherical Symmetries and Static Fields}
Stable ansatze for $U$ minimizing energy and with winding numbers $|n| \geq
2$ have been numerically found by various groups
\cite{brtoca,carson,batsut,homasu}. Energy densities for these static
configurations have been plotted and the remarkable fact has been
discovered that they are invariant under discrete subgroups $G_R$ of the
spatial rotation group $SO(3)_R$. In this section, we briefly explore the
theoretical and possible phenomenological implications of this discovery.

The group $G_R$ is the symmetry group of energy density. It is not
necessarily the invariance group of the static $U$-field. Published work
does not report on the symmetry of the latter.

The two-flavor chiral Lagrangian is invariant under $SO(3)_R \times
SO(3)_I$ where $SO(3)_I$ is the isospin group. The former acts on the
operator $U(x)$ according to $U_R(s_R)U(x)U_R^{-1}(s_R)=U(s_R \vec{x},t)$
while the latter does so according to
$U_I(\gamma_I)U(x)U_I^{-1}(\gamma_I)=\gamma^{\dagger}_I U(x)
\gamma_I$. Here $s_R \in SO(3)_R, \gamma_I \in SU(2)_I$ (the covering group
of $SO(3)_I$) and $U_R(s_R), U_I(\gamma_I)$ are their quantum operators. We
also identify $SO(3)$ and $SU(2)$ with their standard defining matrix
groups.

Suppose that $U_c$ is a static configuration with its energy density having
the symmetry of $G_R$. Then $U_c$ can be invariant under any subgroup of
$SO(3)_R \times SO(3)_I$ which restricted to the first factor is
$G_R$. That is because the energy density being invariant under $SO(3)_I$
will have the symmetry $G_R$.

There can be subgroups $G_I$ of $SO(3)_I$ alone, in addition to subgroups
involving also $G_R$, which leave $U_c$ invariant. There is no need for
$G_I$ too to be discrete.

If ${\cal G}_I \subset SU(2)_I$ is the double cover of $G_I$, its elements
commute with $U_c$. If ${\cal G}_I$ is irreducible, then $U_c$, being in
$SU(2)$, is just $\pm {\bf 1}$ by Schur's lemma. So ${\cal G}_I$ must be
reducible, and hence ${\cal G}_I$ and $G_I$ must be ${\mathbb Z}_N$ or
$U(1)$ groups. We can think of them if necessary after an
$SU(2)$-conjugation to consist of elements of the form $e^{i \tau_3
\xi}$. Excluding the trivial case ${\cal G}_I = \{ \pm {\bf 1} \}$, we can
thus see that $U_c(\vec{x})=e^{i \tau_3 \theta(\vec{x})}$ and that ${\cal
G}_I$ and $G_I$ are in fact $U(1)$.

Let ${\cal G}_R \subset SU(2) \equiv SU(2)_R$ be the inverse image of $G_R$
for the homomorphism $SU(2)_R \rightarrow SO(3)_R$ with $\alpha_R
\rightarrow a_R$. Then there must also exist a subgroup 
\begin{equation}
\hat{\cal G}_R = \{ (\alpha_R, {\cal L}(\alpha_R)): {\cal L} \quad 
\text{a homomorphism of ${\cal G}_R$ into} \quad 
SU(2)_I \}
\end{equation}
of $SU(2)_R \times SU(2)_I$ leaving $U_c$ invariant:
\begin{equation}
{\cal L}(\alpha_R) U_c(a_R^{-1} \vec{x}){\cal L}(\alpha_R)^{-1} =
U_c(\vec{x}).
\end{equation}
Note that if $L(a_R)$ is the image of ${\cal L}(\alpha_R)$ in $SO(3)_I$, we
can write 
\begin{equation}
{\cal L}(\alpha_R) \tau_j {\cal L}(\alpha_R)^{-1} = \tau_k L(a_R)_{kj}.
\end{equation}

The full symmetry group of $U_c$ is 
\begin{equation}
{\cal D} = \hat{\cal G}_R \times \hat{\cal G}_I, 
\end{equation}
where
\begin{equation}
\hat{\cal G}_I = \{ {\bf 1} \} \times {\cal G}_I.
\end{equation} 

We have already found those $U_c$ invariant under $\hat{\cal G}_I$. Let us
now do the same job for $\hat{\cal G}_R$ and ${\cal D}$.

We begin with simple preliminary observations: Any function $\chi(\vec{x})$
has the expansion $\sum c_{lm}(r) Y_{lm}(\hat{x}), \hat{x}=\vec{x}/r$. If
$s_R \in SO(3)$ and we identify $\hat{x}$ with its third column,
$\hat{x}=(s_R)_{i3}$, then also, $Y_{lm}(\hat{x})=\text{constant} \times
D^l_{m0}(s_R)$, where $D^l(s_R)$ are the rotation matrices in the
conventional basis. Hence $\chi(\vec{x})=\sum c_{lm}(r)D^l_{m0}(s_R)$.

Now suppose that ${\cal L}$ is the trivial homomorphism, that is, ${\cal
L}: {\cal G}_R \rightarrow \{ {\bf 1} \}$. In that case, the ansatz for
$U_c$ can be found as follows. Write $U_c(\vec{x})=e^{i \sum \tau_j
\phi_j(\vec{x})}, \phi_j(\vec{x}) \in {\mathbb R}$. Then $\phi_j(a^{-1}_R
\vec{x}) = \phi_j(\vec{x})$. Expanding $\phi_j$,
\begin{equation}
\phi_j(\vec{x})= \sum c^j_{lm}(r)D^l_{m0}(s_R),
\label{expandphi}
\end{equation}
we find that
\begin{equation}
c^j_{lm}(r)=0 \quad \text{unless} \quad D^l_{m0}(a^{-1}_R
s_R)=D^l_{mm'}(a^{-1}_R)D^l_{m' 0}(s_R)\quad \text{is} \quad D^l_{m0}(s_R).
\label{aaction}
\end{equation}
In other words, only those indices $m$ transforming trivially under the
action of $a_R$ [$D^l_{mm'}(a^{-1}_R)=\delta_{mm'}$] survive in
(\ref{expandphi}). 

If ${\cal L}$ is non-trivial, we still have the expansion
(\ref{expandphi}), but the conditions on $c^j_{lm}$ are different, being
\begin{equation}
[\tau_k L(a_R)_{kj}][c^j_{lm}(r)D^l_{m0}(a^{-1}_R s_R)]=\tau_k
c^k_{lm}(r)D^l_{m0}(s_R).
\end{equation}
Let $\bar{m}, \bar{m}'$.. denote indices transforming by the representation
$L$ of ${\cal G}_R$:
\begin{eqnarray}
{\cal L}(\alpha_R)\tau_{\bar{m}}{\cal L}(\alpha_R)^{-1} &=& 
\tau_{\bar{m}'}L(a_R)_{\bar{m}' \bar{m}}, \\
D^l_{\bar{m}0}(a^{-1}_R s_R) &=& L(a^{-1}_R)_{\bar{m} \bar{m}'}
D^l_{\bar{m}'0}(s_R).
\end{eqnarray}
There could be such indices for several values of $l$. Even for one $l$,
the representation $L$ could occur with multiplicity, so we can write
$(\bar{m}, \nu)$, where $\nu$ accounts for this multiplicity. The
requirement of ${\cal G}_R$-invariance is now seen to be accounted for by
writing 
\begin{eqnarray}
U_c(\vec{x})&=&e^{i \sum_{\bar{m}} \tau_{\bar{m}} \phi_{\bar{m}}(\vec{x})},
                \label{Utrn}\\ 
\phi_{\bar{m}}(\vec{x})&=&\sum_{l, \nu} a_{l \nu}(r) D^l_{(\bar{m}, \nu),
                0}(s_R) .
\end{eqnarray}

This formula is perfectly general if $L$ is irreducible, that is if $G_R
\neq {\mathbb Z}_N$ or $D_2$. If $G_R$ is ${\mathbb Z}_N$, then $L$ splits
into three irreducible representations (IRR's), the image of the generator
$z$ of $G_R$ in these IRR's being $e^{2\pi i/N}, e^{-2\pi i/N}$ and 1. The
index $\bar{m}$ can be taken to be $+, - $ and 3 for these IRR's. With
$\tau_{\pm}=\tau_1 \pm i \tau_2$, we now have the freedom to generalize
(\ref{Utrn}) to
\begin{equation}
U_c(\vec{x})=e^{i \{ [c_{+}\tau_{+}\phi_{+}(\vec{x}) + \text{complex
conjugate}] + c_3 \tau_3 \phi_3(\vec{x}) \} }, \quad c^{*}_3=c_3.
\end{equation}
A similar generalization exists for $D_2$.

There is a difference between $\hat{\cal G}_R$ and ${\cal D}$ if ${\cal
G}_I=U(1)$. But then $U_c(\vec{x})=e^{i \tau_3 \theta(\vec{x})}$, and
$L(a_R)$ can only be a ${\mathbb Z}_N \subset U(1)$. As it can be absorbed
in $U(1)$, we are just in a special case of having a trivial $L$.

\subsection{Collective Coordinates}
As the symmetry of the Hamiltonian is $SO(3)_R \times SO(3)_I$, collective
coordinates are introduced using its action on $U_c$, that is by writing 
\begin{equation}
U(x)=\gamma_I(t)U_c[T_R(t)^{-1}\vec{x}]\gamma_I(t)^{-1}, \; T_R(t) \in
SO(3)_R, \gamma_I(t) \in SU(2)_I
\end{equation}
and quantizing $T_R$ and $\gamma_I$. The actions 
\begin{eqnarray}
\hat{\cal G}_R \ni (\alpha_R, {\cal L}(\alpha_R))&:&(T_R, \gamma_I)
       \rightarrow (T_R a_R, \gamma_I {\cal L}(\alpha_R)), \\
\hat{\cal G}_I \ni ({\{\bf 1}, \beta_I)&:&(T_R, \gamma_I) \rightarrow (T_R,
       \gamma_I \beta_I) 
\end{eqnarray}
of the symmetry group ${\cal D}$ leaves $U$ invariant.

The eigenfunctions $\;\psi\;$ of the $\;$ Skyrme $\;$Hamiltonian are
functions on \\ $SU(2)_R \times SU(2)_I$, the universal cover of $SO(3)_R
\times SO(3)_I$. We can denote its coordinates by $(\tau_R, \gamma_I)$, and
identify $T_R$ as the image of $\tau_R$ under the homomorphism $SU(2)_R
\rightarrow SO(3)_R$ \cite{BMSS,giulini}. 

The group ${\cal D}$ acts on $(\tau_R, \gamma_I)$,
\begin{eqnarray}
(\tau_R, \gamma_I) \rightarrow (\tau_R \alpha_R, \gamma_I{\cal
        L}(\alpha_R)),\\
(\tau_R, \gamma_I) \rightarrow (\tau_R, \gamma_I \beta_I)
\end{eqnarray}
and hence also on $\psi$. A basic result \cite{BMSS,giulini} is that
$\psi$'s are vector-valued, $\psi = (\psi_m)$ and transform on the index
$m$ by an irreducible representation $\rho$ of ${\cal D}$:
\begin{eqnarray}
\psi_m(\tau_R \alpha_R, \gamma_I {\cal L}(\alpha_R)) &=& \psi_n(\tau_R,
    \gamma_I) \rho_{nm}(\alpha_R, {\cal L}(\alpha_R)), \label{Ponpsi1} \\
\psi_m(\tau_R, \gamma_I \beta_I) &=& \psi_n(\tau_R, \gamma_I)
    \rho_{nm}(({\{\bf 1}, \beta_I)) \label{Ponpsi2}.
\end{eqnarray}
There are also restrictions on $\rho$ from the underlying quark model:
$\psi$ must be spinorial under {\it both} spin and isospin rotations, or it
must be tensorial under both.

\subsection{What Parity Does to Symmetries}
An adaptation of our previous work \cite{basiwi,balvai1,balvai2} to the
present situation shows that the parity transform ${\cal P}U_c$ of $U_c$
can be obtained by applying a particular element $(s_P, \iota_P)$ of
$SO(3)_R \times SU(2)_I$:
\begin{equation}
({\cal P}U_c)(\vec{x}) = U_c(-\vec{x})^{\dagger}=\iota_P
U_c(s^{-1}_P\vec{x}) \iota^{-1}_P. 
\label{PonU}
\end{equation}
That is because the parity transform of the energy density can be obtained
by applying a spatial rotation.

Let $\zeta_P$ be an inverse image of $s_P$ in $SU(2)$. It is uncertain upto
a sign, so we pick one and call it $\zeta_P$. The symmetry group of ${\cal
P}U_c$ is then $(\zeta^{-1}_P, \iota^{-1}_P){\cal D}(\zeta_P, \iota_P)$.
Also as ${\cal P}$ commutes with $SO(3)_R \times SO(3)_I$, we have the
fundamental result
\begin{equation}
(\zeta^{-1}_P, \iota^{-1}_P){\cal D}(\zeta_P, \iota_P) = {\cal D}.
\end{equation}

The parity transform ${\cal P}\psi$ of $\psi$ is specified by 
\begin{equation}
({\cal P}\psi)_m(\tau_R, \gamma_I)= \psi_m(\tau_R \zeta_P, \gamma_I
\iota_P).
\end{equation}

Its transformation under ${\cal D}$ is given, according to
(\ref{Ponpsi1},\ref{Ponpsi2}) by 
\begin{eqnarray}
({\cal P}\psi)_m(\tau_R \alpha_R,\gamma_I{\cal L}(\alpha_R))&=&\nonumber \\
 ({\cal P}\psi)_n(\tau_R, \gamma_I)\!\!\!\!&\rho&\!\!\!\!\!_{nm}(\zeta^{-1}_P
 \alpha_R \zeta_P, \iota^{-1}_P {\cal L}(\alpha_R) \iota_P) \\ 
({\cal P}\psi)_m(\tau_R, \gamma_I \beta_I) &=& ({\cal P}\psi)_n 
(\tau_R, \gamma_I) \rho_{nm}({\{\bf 1}, \iota^{-1}_P \beta_I \iota_P).
\end{eqnarray}

The $\rho$ with its arguments here defines the parity transform ${\cal
P}\rho$ of the representation $\rho$:
\begin{eqnarray}
{\cal P}\rho: (\alpha_R, {\cal L}(\alpha_R)) &\rightarrow&
	      \rho(\zeta^{-1}_P \alpha_R \zeta_P, \iota^{-1}_P {\cal
	      L}(\alpha_R) \iota_P), \\
	      ({\{\bf 1}, \beta_I) &\rightarrow& \rho({\{\bf 1},
	      \iota^{-1}_P \beta_I \iota_P).
\end{eqnarray}
It can happen that ${\cal P}\rho$ is inequivalent to $\rho$, ${\cal P}\rho
\neq \rho$. That can be the case for example when $G_R$ is a dihedral or a
${\mathbb Z}_N$ group \cite{basiwi,balvai1,balvai2}. When that is so, our
previous work predicts a pair of approximately degenerate states.

Actually when ${\cal P}\rho \neq \rho$, what is predicted in the absence of
quarks is a breakdown of parity by a mechanism like that of the QCD
$\theta$-angle. But with quarks, parity is restored, but we are left with
parity doubles.

In a similar manner, we can establish the existence of time-reversal doubles
if $\rho$ is a complex representation. Time reversal ${\cal T}$ is actually
broken in the absence of quarks, but is restored when quarks are included. 

In molecular physics, it is the electrons which restore the ${\cal P}$-
and ${\cal T}$- symmetries ruined by the underlying quantum shapes. The
role of electrons is thus assumed here by quarks.

The mechanism restoring $\cal P$ and $\cal T$ via the B-O approximation is
the same as that discussed in the last section. Let us recall it, adapting
it appropriately. By assumption, the domain $V^{(\rho_0)}$ of the full
Hamiltonian ${\cal H}$ must be associated with the trivial representation
${\rho}_0$. An eigenstate ${\psi}^{({\overline{\rho}})}_f$ of ${\cal H}_f$
is a section of a vector bundle over $Q$ in the B-O approximation (the
superscripts on the wave functions indicate the UIR).  When the B-O
Hamiltonian ${\hat{\cal H}}_s$ is calculated as discussed in the previous
section, it can be shown to contain a connection and has a domain
associated with the UIR $\rho$, the complex conjugate of
$\overline{\rho}$. So an eigenstate ${\psi}^{({\rho})}_s$ of ${\hat{\cal
H}}_s$ corresponds to $\rho$ and the product wave function
${\psi}={\psi}^{({\rho})}_s\; {\psi}^{({\overline{\rho}})}_f$ corresponds
to $\rho \otimes \overline{\rho}$. But $\cal H$ and ${\cal H}_s$ act on the
total wave function and their domain can only correspond to ${\rho}_0$.That
is now easily arranged as ${\rho}_0$ occurs in the reduction of $\rho
\otimes \overline{\rho}$. The correct total wave function in the B-O
approximation is thus the orthogonal projection ${\chi}^{({\rho}_0)} = {\bf
P}[{\psi}^{({\rho})}_s \;{\psi}^{({\overline {\rho}})}_f]$ of $\psi$ to
$V^{({\rho}_0)}$.  Now it can happen that the parity transform
${\cal P}\rho$ of $\rho$ is $\overline {\rho}$ and hence that of
$\overline {\rho}$ is $\rho$. The parity transform ${\cal P}
{\chi}^{({\rho}_0)}$ of ${\chi}^{({\rho}_0)}$ is then of the form ${\bf
P}[{\psi}^{({\overline{\rho}})}_s \;{\psi}^{({\rho})}_f] \in
V^{({\rho}_0)}$. It is still in the domain of $\cal H$ and ${\cal H}_s$, so
there is no question of $\cal P$-violation. The same goes for $\cal T$. But
there is a doubling of states. The doubles with definite $\cal P$, for
example, in the leading approximation are linear combinations of
${\chi}^{({\rho}_0)}$ and ${\cal P}{\chi}^{({\rho}_0)}$.

In this manner, we can see that when quantum Skyrmions violate $\cal P$ or
$\cal T$, then we may have $\cal P$ or $\cal T$ doubles and no $\cal P$ or
$\cal T$ symmetry breakdown after fast variables (in this case the quarks)
are included.

Generally, some higher order interactions will lift the degeneracy between
these parity doublets, and so what we expect to see experimentally are
pairs of approximately degenerate states with opposite parity, all other
quantum numbers (except energy) being identical. A candidate mechanism for
lifting this degeneracy was discussed in \cite{balvai2}.

We can now ask if any of the light nuclei exhibit any parity doubling in
the low energy spectra. The spectra of $^9Be$ and $^9B$ show clear evidence
of such doubling \cite{selove}. There are at least three low-lying pairs of
states with all quantum numbers identical except parity (and energy) for
$^9Be$. Similarly, there are at least two such pairs for $^9B$. The typical
separation between parity partners is of the order of 0.5 to 5 MeV. The
$B=9$ Skyrmion has a tetrahedral symmetry, and may thus be a good candidate
for illustrating the mechanism discussed in this paper.

\section{Conclusions and Outlook}
Using spectral flow arguments, we have shown that we can strengthen the
identification \cite{bnrs1,bnrs2} of the topological winding number of the
Skyrmion with the fermion number. In addition, our arguments naturally lead
to the existence of quarks moving about in the field of this Skyrmion. The
parity-even combination of the quark states correctly describes the nucleon
and delta, whereas the parity-odd combination describes states $I (J^P) =
\frac{1}{2} (\frac{1}{2}^-)$ and their excitations. There are candidates
for such states in the Particle Data booklet, such as $N(1535)$ and 
$N(1650)$. Using the B-O approximation, we also showed that when the fast
quarks are averaged out, the Skyrmion itself becomes spinorial. As
discussed in \cite{balvai2} (where it was argued that systems with certain
discrete symmetries must exhibit a parity doublet structure), we also argue
that some nuclei may show such a structure in their low energy spectra. We
suggest the case of the $B=9$ nuclei to support our claim.

Our focus has been on the case of two flavors. To discuss $N_f \geq 3$, we
start by looking at a generalized spherically symmetric ansatz that is
analogous to the hedgehog ansatz (See for example, Ch.17 of \cite{BMSS}).
The B-O treatment of this situation involves non-flat bundles. As discussed
in \cite{balvai2}, the application of our methods to this case is
straightforward.

\section{Acknowledgments}
We would like to thank E. Dambasuran for discussions during early stages of
this work, P. Golumbeanu for valuable discussions and partial collaboration
during the work and P. M. Sutcliffe for clarifications regarding
\cite{homasu}.  This work was supported in part by the US DOE under
contract number DE-FG02-85ER40231.

\bibliographystyle{unsrt}

\bibliography{skyrmion}

\end{document}